\documentclass[12pt,preprint]{aastex}

\slugcomment{}

\shorttitle{\textit{AKARI} Observations of Stephan's Quintet}
\shortauthors{Suzuki et al.}

\begin{document}


\title{Far-infrared emission from intergalactic medium in Stephan's
   Quintet revealed by \textit{AKARI}}


\author{Toyoaki Suzuki\altaffilmark{1}, Hidehiro Kaneda\altaffilmark{2},
Takashi Onaka\altaffilmark{3}, \and Tetsu Kitayama\altaffilmark{4}}
\email{suzuki@ir.isas.jaxa.jp}

\altaffiltext{1}{Institute of Space and Astronautical Science, Japan Aerospace
   Exploration Agency, 3--1--1 Yoshinodai, Chuo-ku,
   Sagamihara, Kanagawa 252--5210, Japan}
\altaffiltext{2}{Graduate School of Science, Nagoya University,
   Furu-cho, Chikusa-ku, Nagoya 464--8602, Japan} 
\altaffiltext{3}{Department of Astronomy, Graduate School of Science, The
University of Tokyo, 7-3-1 Hongo, Bunkyo-ku, Tokyo 113-0033, Japan}
\altaffiltext{4}{Department of Physics, Toho University, Funabashi,
   Chiba 274-8510, Japan}


\begin{abstract}
 The Stephan's Quintet (SQ, HCG92) was observed with the Far-Infrared Surveyor
 (FIS) aboard \textit{AKARI} in four far-infrared~(IR) bands at 65, 90,
 140, and 160 $\mu$m. The \textit{AKARI} four-band images of the SQ show
 far-IR emission in the intergalactic medium (IGM) of the SQ. In
 particular, the 160 $\mu$m band image shows single peak emission in
 addition to the structure extending in the North-South direction along the shock ridge as seen in 
 the 140 $\mu$m band, H$_2$ emission and X-ray emission.  
 Whereas most of the far-IR emission in the shocked region comes from 
 the cold dust component, shock-powered [\ion{C}{2}]158\,$\mu$m emission
 can significantly contribute to the emission in the 160
 $\mu$m band that shows a single peak at the shocked region. In the
 shocked region, the observed gas-to-dust mass ratio is in agreement
 with the Galactic one. The color temperature of the cold dust component
 ($\sim$20 K) is lower than that in surrounding galaxies ($\sim$30
 K). We discuss a possible origin of the intergalactic dust emission.  
\end{abstract}


\section{Introduction}
\begin{figure}
\begin{center}
 \includegraphics[angle=0,scale=0.40]{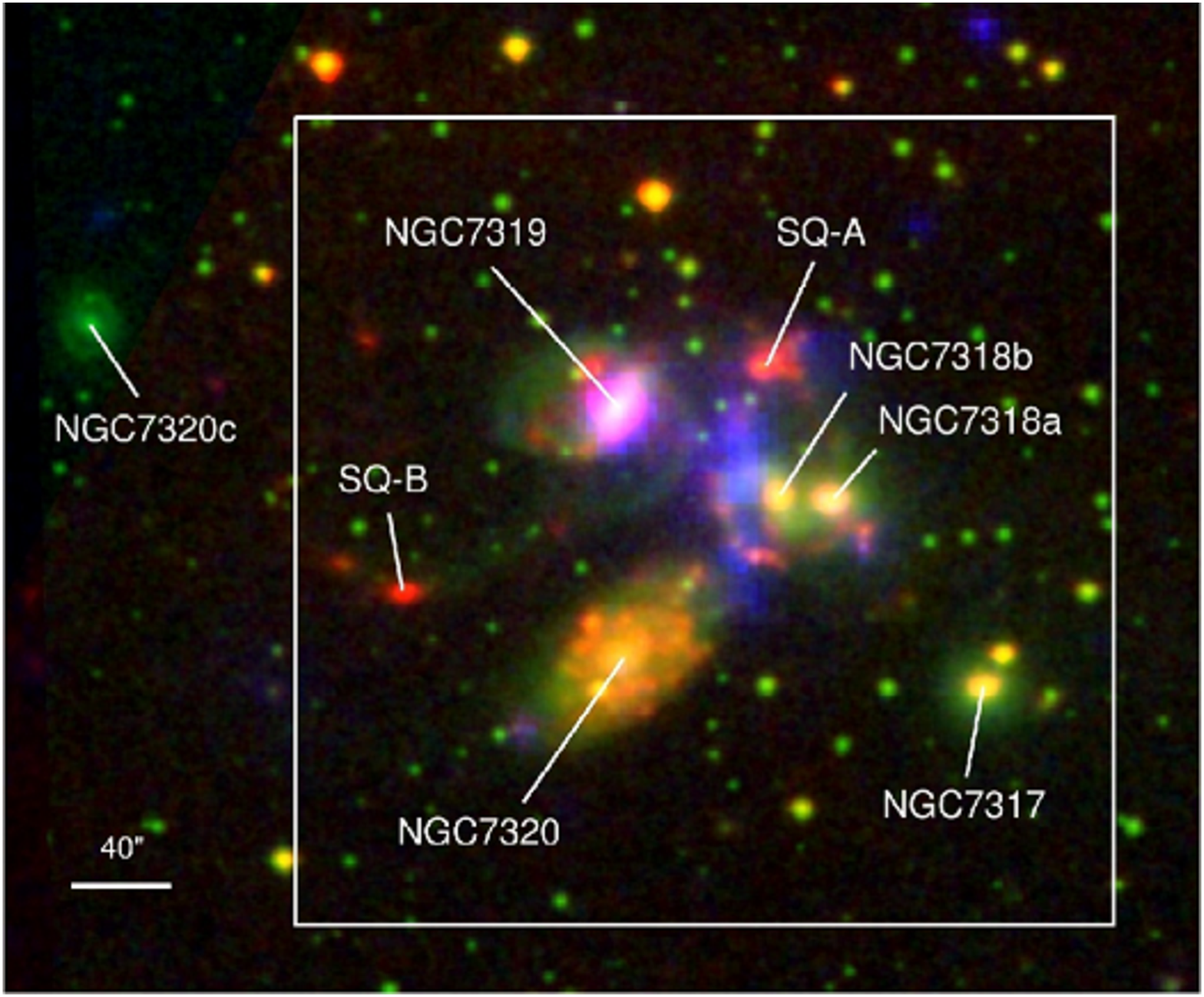}
\end{center}
 \caption{Composite image of the Stephan's Quintet (HCG92): \textit{AKARI} 7
 $\mu$m (red), Optical (green), and X-ray (blue). Optical and X-ray
 images are retrieved from Digitized Sky Survey and XMM-Newton
 Science Archive (observation ID: 0021140201, observation date:
 2001-12-07), respectively. The SQ is composed of five galaxies (NGC
 7317, NGC 7318a, NGC 7318b, NGC 7319, and NGC 7320c). NGC 7320 is an 
 unrelated foreground object~\citep{moles1997}.  The box shows the
 boundary of the maps in Fig.~\ref{4image}. } 
 \label{SQ}
\end{figure}

The track of the galactic scale transmigration of gas and dust 
is indispensable to understand the galaxy evolution. 
Stephan's Quintet (SQ, HCG92) is the well studied compact group of
galaxies~\citep{hickson} with the disturbed intergalactic medium~(IGM) as shown in 
Fig.~\ref{SQ}. The intruder
galaxy NGC 7318b is currently colliding with the IGM at a relative
velocity of $\sim1000$ km s$^{-1}$ and causes a
large-scale shock front and IGM starbursts (SQ-A and
SQ-B). The shock front was first discovered
by~\cite{allen1972} with radio, and
subsequently detected in the X-ray emission from shock-heated
($\sim6\times 10^6$ K) gas~\citep{pietsch, trinchieri2003,
trinchieri2005, osullivan}. 
 \cite{appleton} found powerful H$_2$ rotational line emission from
warm ($\sim 10^2$-$10^3$ K) molecular gas in the center of the shock
ridge. The significant discovery is the extremely large equivalent width
($EW$) of the H$_2$ line emission. This may indicate that far-infrared (IR)
fine-structure lines such as [\ion{C}{2}]158\,$\mu$m line also show
extremely large $EW$ by the shock and have a significant contribution to far-IR
luminosity. However, the fact has yet to be revealed observationally.    
\cite{culver2010} made H$_2$ maps of the SQ and found
that the spatial distribution of the H$_2$ emission was similar to those
of radio and X-ray emission in the shocked region. 
To explain the co-existence of
both hot and H$_2$ gas, \cite{guillard2009} proposed a
model of the shock in the inhomogeneous gas medium.
While H$_2$ molecules are likely to be produced on the
grain surface~\citep{gould1963}, direct association of the H$_2$ gas with dust
has yet to be found in the shock ridge. Recently, \cite{natale2010}
reported dust emission in the SQ with \textit{Spitzer} observations.
While the map at 160 $\mu$m shows the significant presence of cold dust
in the shock ridge, it is difficult to make direct comparison of the
diffuse 160 $\mu$m map with the H$_2$ emission map because of the effect
of bright emission from the surrounding galaxies. 

The Far-Infrared Surveyor (FIS)~\citep{kawada} aboard
\textit{AKARI}~\citep{murakami} has four far-IR bands with the central 
wavelengths of 65, 90, 140, and 160 $\mu$m, and achieves high
sensitivity with relatively high spatial
resolution. Finer allocation of \textit{AKARI}/FIS 4
bands can provide the spectral information to constrain both the
dust temperature and the [\ion{C}{2}]158\,$\mu$m line emission and is
the distinctive advantage compared to \textit{Spitzer}/MIPS observations. 
In this paper we report the observation of the SQ with the FIS, and
discuss the origin of the far-IR emission in the IGM.

\section{Observations and Data reduction}
The SQ was observed in part of the \textit{AKARI} mission program ``ISM in our
Galaxy and Nearby Galaxies'' \citep[ISMGN;][]{kaneda2009}  on 2007 Jun 18
using the FIS01 observation mode (Observation ID: 1402238-001). Details of
the FIS instrument and its in-orbit performance/calibration are
described in~\citet{kawada} and \citet{shirahata}.     
The FIS was operated in the photometry mode with the four bands: N60
(65 $\mu$m), WIDE-S (90 $\mu$m), WIDE-L (140 $\mu$m),
and N160 (160 $\mu$m). The four-band data were simultaneously
obtained in a pointing observation for an area of about
$10\arcmin\times20\arcmin$ around the SQ. The FIS data were processed
with the \textit{AKARI} official pipeline modules (version 20070914).
Finally, four-band
images were created with grid sizes of $12.5\arcsec$ for the
WIDE-L and N160 bands and $7.5\arcsec$ for the
WIDE-S and N60 bands. The widths (FWHM) of the point
spread functions (PSFs) are $\sim45\arcsec$ for the WIDE-L and
N160 bands and $\sim30\arcsec$ for the WIDE-S and
N60 bands \citep{shirahata}. The four-band images are
presented in Fig.~\ref{4image}. The images are smoothed with boxcar
kernels with a width of $25\arcsec$ for the WIDE-L and
N160 bands, and $15\arcsec$ for the WIDE-S and
N60 bands.

To derive the spatial variation in color temperature,  the spatial
resolutions of the original WIDE-S and N60 images are
reduced to match those of the WIDE-L and N160 images
by convolving the former images with a Gaussian kernel. The images are then
resized with the common spatial scale among the four bands,
$12.5\arcsec$/pixel. The four-band flux densities at each image bin are 
derived after subtracting the sky background. The background
levels were estimated and subtracted using nearby regions of the blank sky,
which were observed in the beginning and at the end of the scan. Color
corrections were also applied for the obtained flux densities by
assuming a modified blackbody spectrum with the emissivity power-law index
of unity and the temperatures of 20 K and 30 K for the shocked region and
galaxies, respectively. Aperture corrections were not applied to the flux densities
because we only consider diffuse emission components. At the centers of
the shocked region, NGC 7319, and NGC 7320, we derive the flux
densities by integrating the surface brightness within circular
apertures. Figure~\ref{sed}(a) shows the resulting spectral energy distributions 
(SEDs) at these positions. Note that the fluxes in Fig.~\ref{sed}(a) are
not the fluxes for the whole areas of the shocked region and the galaxies.
The systematic errors for the flux densities
are estimated to be $\sim$10 \% for the WIDE-S and N60 bands, 
$\sim$20 \% for the WIDE-L band, and $\sim$25 \% for N160 band, which
include the uncertainty due to radiation effects. The relative
uncertainties for the flux densities are estimated to be below 5 \% for
the four bands.

\section{Results}
\begin{figure}
 \includegraphics[width=18cm]{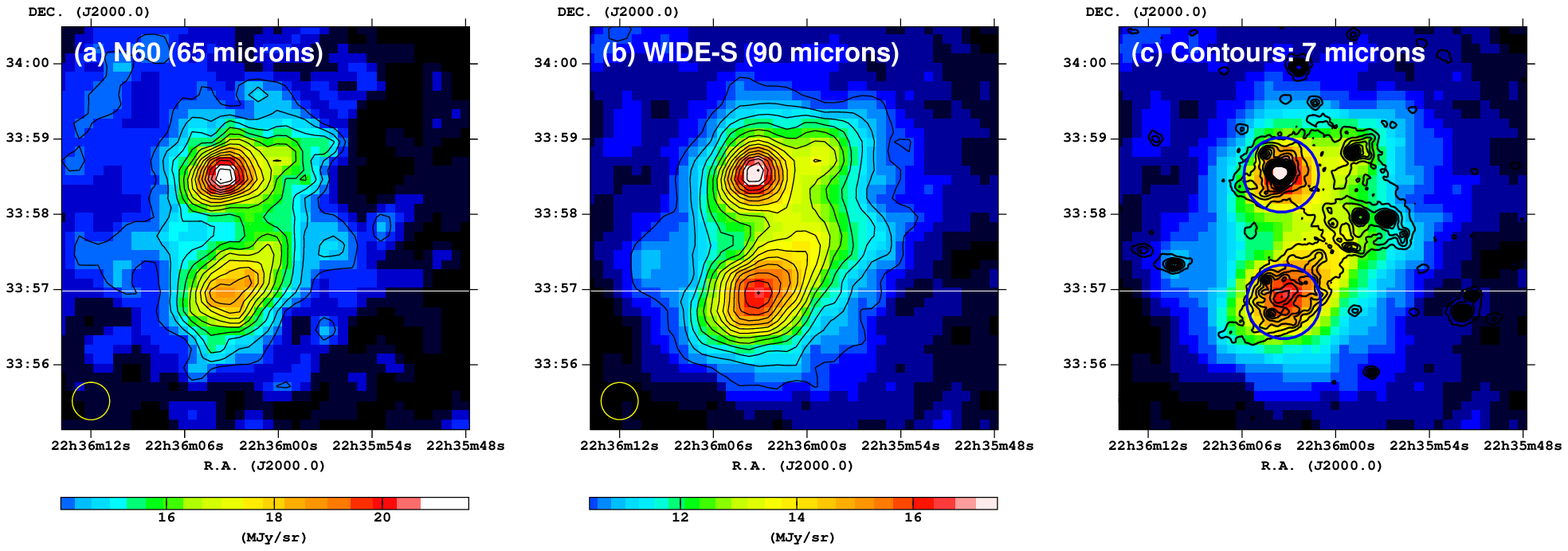}
 \includegraphics[width=18cm]{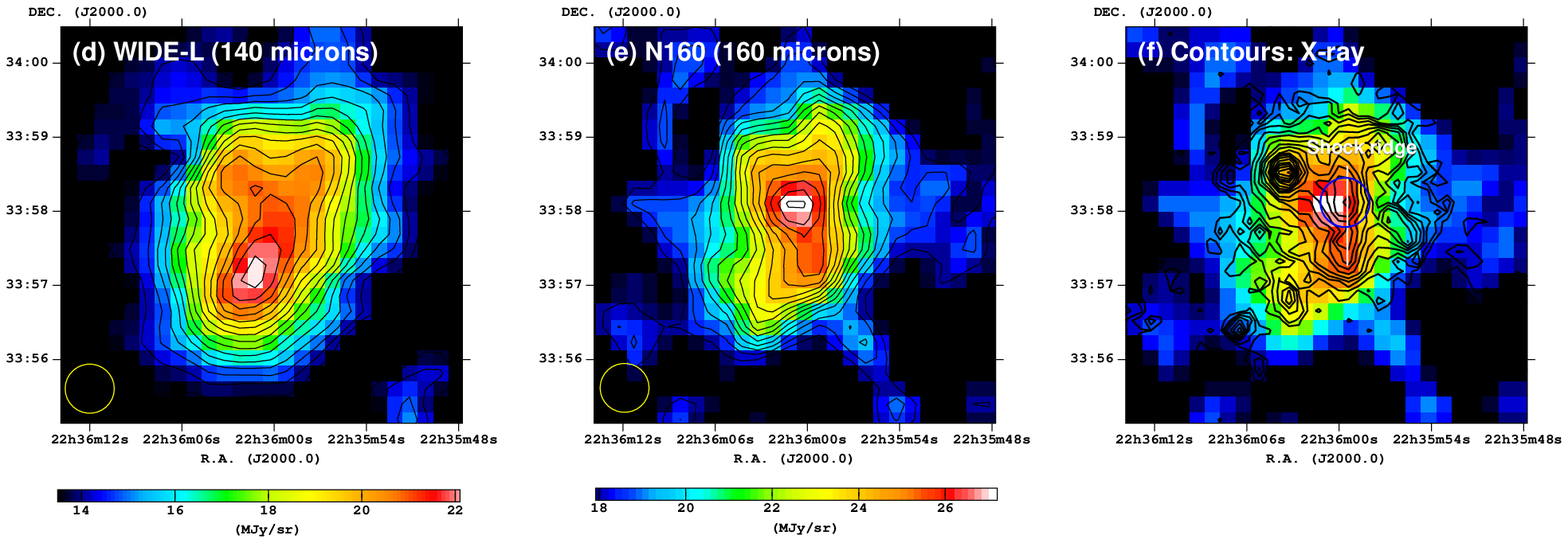}
 \caption{Four-band images of the SQ in the N60 (top-left),
 WIDE-S (top-middle), WIDE-L (bottom-left), and
 N160 (bottom-middle) bands. The center wavelengths of the four
 bands are 65 $\mu$m for N60, 90 $\mu$m 
 for WIDE-S, 140 $\mu$m for WIDE-L, and 160 $\mu$m for
 N160. The contours are linearly spaced from 10 \% to 99 \% of
 the peak brightness with a step of 3 \%. The peak brightness that
 includes the sky background is 22 MJy sr$^{-1}$ (N60), 17 MJy sr$^{-1}$
 (WIDE-S), 22 MJy sr$^{-1}$ (WIDE-L), and 27 MJy sr$^{-1}$ (N160). In
 each image, the PSF size in FWHM is shown in the lower left corner. The
 upper-right and bottom-right panels show the images of WIDE-S and N160
 bands overlaid on \textit{AKARI} 7 $\mu$m 
 and XMM/Newton X-ray contours, respectively. The three apertures for
 photometry of the shocked region ($r=20''$), NGC 7319 ($r=30''$), and
 NGC 7320 ($r=30''$) are shown in the panels of (c) and (f). As a
 reference, the shock ridge is shown by the solid white line in the panel (f).}
 \label{4image}
\end{figure}

\begin{figure*}[t]
\vspace{-40mm}
  \begin{center}
   \hspace{-33mm}
    \includegraphics[angle=0,scale=.50]{fig3a.ps}
  \end{center}
\vspace{-3mm}
  \begin{center}
   \hspace{-20mm}
    \includegraphics[angle=0,scale=.65]{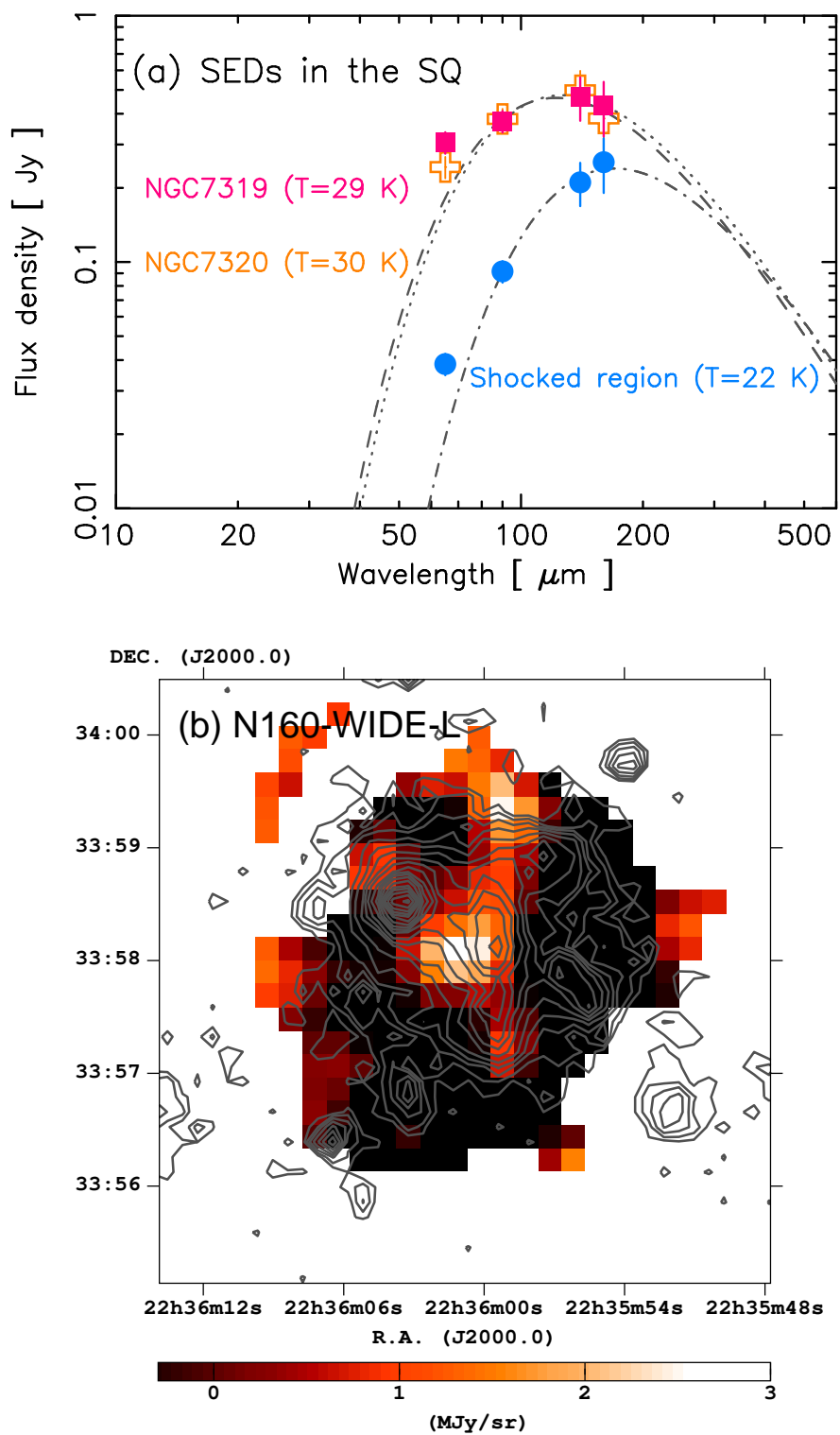}
  \end{center}
 \vspace{-5mm}
 \caption{(a) SEDs at the shocked region, NGC 7319, and
 NGC 7320. The positions of the apertures are shown in Figs~\ref{4image}(c)
 and (f). The dashed, dotted, and dash-dotted lines show the single
 temperature modified blackbody spectrum with the emissivity power-law
 index of unity for NGC 7319, NGC 7320, and the shocked region,
 respectively. 
(b) The N160 image shown after subtraction of the WIDE-L 
image. The contours show XMM/Newton X-ray
 contours that are the same as those in Fig.~\ref{4image}(f). }    
\label{sed}
\end{figure*}

In Fig.~\ref{4image}, N60 and WIDE-S images
show emission from NGC 7319 and NGC 7320, 
and from the SQ-A and SQ-B (Fig.~\ref{4image}(c)). 
Although the WIDE-L image shows a structure extending in the North-South direction
along the shock ridge, the N160 image clearly shows single peak emission
in addition to the structure, but does not indicate any features
associated with the individual galaxies. In the shocked region, the
background-subtracted surface brightness in the N160 band is larger than
9 MJy sr$^{-1}$, which is comparable to 5$\sigma$ sky background
noise. For the first time, \textit{AKARI} observations reveal the
dramatical change in the spatial distribution of far-IR emission at
wavelengths around 140-160 $\mu$m.    
The difference in images between the N160 band and
the MIPS160 band~\citep{natale2010} may come from the fact that the
spectral response of the 
N160 band covers longer wavelengths than that of the MIPS160
band; wavelength coverage at 10 \% of the
peak responsivity is $140$-$196$ $\mu$m for the N160 band~(AKARI FIS
Data Users Manual ver.1.3) and $129$-$184$ $\mu$m for the MIPS160
band~(MIPS Instrument Handbook). Figure~\ref{4image}(f) shows that the
spatial distribution of 160 $\mu$m emission is quite similar to that of X-ray emission in the
shocked region. In fact, X-ray, radio, and H$_2$ emission in the shocked
region all show structures associated with the shock ridge and the
bridge extending eastward from the shock ridge to NGC 7319. Our result
clearly shows the spatial correlation of the far-IR emission with these
structures.

\subsection{Difference in images between N160 and
  WIDE-L bands}
The cold dust temperatures in the shock and surrounding galaxies are
about 20-30 K. Because the flux ratio between the N160 and
WIDE-L bands is not very sensitive to the dust temperature, 
the difference in the spatial distribution of far-IR emission between
the two bands is hard to explain only by the presence of the cold
dust component. An alternative possibility is a contribution from the shock-powered
[\ion{C}{2}]158\,$\mu$m line emission to the N160 band.  
For the SQ, the wavelength of [\ion{C}{2}]158\,$\mu$m line is redshifted
to 161 $\mu$m~\citep{hickson1992}. Thus, its contribution to the 
N160 band is larger than to the WIDE-L band. 
To estimate the contribution from the [\ion{C}{2}] line, the WIDE-L flux
density $F_\mathrm{WL}$ is simply subtracted from the N160 flux
density $F_\mathrm{N160}$ by assuming that
the dust emission is constant in units of $F_\nu$ over the two bands
(Fig.~\ref{sed}(b)). By using the same aperture as that in
Fig.~\ref{4image}(f), the subtracted flux density $F_\mathrm{sub}$ is
estimated to be 40$^{+18}_{-22}$ mJy. The error includes the uncertainty
that $F_\mathrm{WL}$ to be subtracted must be increased by 38 \%,
if the SED of the dust emission is considered as the modified blackbody
spectrum whose temperature and amplitude are determined from the data at
90 and 140 $\mu$m. The uncertainty is comparable to the subtracted sky
background noise. The [\ion{C}{2}] luminosity surface density
$\Sigma_{L_{\mathrm{[CII]}}}$~[erg sec$^{-1}$ kpc$^{-2}$] is given by
\begin{equation}
 \Sigma_{L_{\mathrm{[CII]}}} = 4\pi D^2
  \frac{F_\mathrm{sub}}{A}\left(\frac{R_\mathrm{N160}(161\mu
		 \mathrm{m})}{\Delta\nu_{\mathrm{N160}}}-\frac{R_\mathrm{WL}(161\mu \mathrm{m})}{\Delta\nu_{\mathrm{WL}}}\right)^{-1},
\label{eq5}
\end{equation}
where $D$ is the distance to the SQ of 94 Mpc, $A$ is the aperture area
		 at the shocked region (260 kpc$^2$), $R_\mathrm{N160}$  
		 and $R_\mathrm{WL}$ are 
		 the relative responses of the N160 and WIDE-L bands
		 at 161 $\mu$m, respectively, and
		 $\Delta\nu_{\mathrm{N160}}$ and $\Delta\nu_{\mathrm{WL}}$ are the
		 effective bandwidths of the N160 and WIDE-L, respectively. From \cite{kawada},
		 $R_\mathrm{N160}$, $R_\mathrm{WL}$, $\Delta\nu_{\mathrm{N160}}$, and
		 $\Delta\nu_{\mathrm{WL}}$ are taken as 0.96, 0.59, 0.4 THz, and
		 0.8 THz, respectively. Thus,
		 $\Sigma_{L_{\mathrm{[CII]}}}$ is estimated to be
		 $(1.0^{+0.4}_{-0.5})\times10^{39}$ erg sec$^{-1}$
		 kpc$^{-2}$, which is comparable to the H$_2$ line
		 luminosity surface density $\Sigma_{L_\mathrm{H_2}}$ of $2\times10^{39}$ erg
		 sec$^{-1}$ kpc$^{-2}$\citep{culver2010}. By using the
		 dust emission at 160 $\mu$m, the $EW$ of the [\ion{C}{2}] line is
		 estimated to be $\sim$10 $\mu$m. The $EW$ is about 10 times larger
		 than that in nearby galaxies (the mean of 0.7 $\mu$m with the dispersion
		 of 0.2-3.0 $\mu$m, \cite{boselli2002}). 
		 We calculated the contribution of the [\ion{C}{2}] line emission to the
		 fluxes in the WIDE-L and N160 bands by considering
		 the constant dust emission in units of $F_\nu$ over the two bands. We
		 obtained 8 \% to WIDE-L band and 23 \% to N160 band.

\subsection{Properties of far-IR dust emission}
To estimate the far-IR luminosity $L_\mathrm{FIR}$ in
the shocked region, the best-fit modified blackbody (the dash dotted line
in Fig~\ref{sed}(a)) was integrated between 3 and 3000 $\mu$m to yield
$L_\mathrm{FIR}$ of $(5 \pm 2)\times 10^{42}$ erg sec$^{-1}$.  
By dividing $L_\mathrm{FIR}$ by the aperture area $A$, the luminosity
surface density $\Sigma_{L_\mathrm{FIR}}$ is estimated to be $(2.0 \pm
0.6)\times 10^{40}$ erg sec$^{-1}$ kpc$^{-2}$, which is in agreement
with the far-IR luminosity surface density ($1.6\times 10^{40}$ erg
sec$^{-1}$ kpc$^{-2}$) of \textit{ISO} observations~\citep{xu2003}.  
In the shocked region, we calculate the dust mass according to Eq.~(4) of
\cite{hildebrand} with the emissivity power-law index of unity.
The dust mass surface density $\Sigma_{M_\mathrm{d}}$ is given by 
\begin{equation}
 \Sigma_{M_\mathrm{d}} = 1.1\times10^4\left(
				    \frac{\Sigma_{L_\mathrm{FIR}}}{2.0\times10^{40}\
			   \mathrm{erg\ sec^{-1}}\ \mathrm{kpc}^{-2}}\right)
 \left(\frac{T_{\rm d}}{22\ \mathrm{K}}\right)^{-5}\ M_\sun\ \mathrm{ kpc^{-2}}. 
\label{eq6}
\end{equation}
The dust temperature $T_\mathrm{d}$ is set to
be 22 K, which is derived from the SED fitting in Fig~\ref{sed}(a).
From Eq.~(\ref{eq6}), $\Sigma_{M_\mathrm{d}}$ is estimated to be $(1.1\pm0.3)\times10^4$
$M_\sun$~kpc$^{-2}$. If we consider the contribution of the [\ion{C}{2}]
line emission to $F_\mathrm{N160}$, $\Sigma_{M_\mathrm{d}}$
is reduced by $\sim$20 \%. \cite{culver2010} estimated the mass surface
density of warm H$_2$ gas in the shocked region as
$(10\pm2)\times10^5$ $M_\sun$~kpc$^{-2}$.    
CO observations of the shocked region suggest that the mass ratio of the
cold H$_2$ ($<$ 50 K) to warm H$_2$ gas masses is 2 with the CO-to-H$_2$ conversion
factor of our Galaxy~\citep{guillard2010b}.
\cite{guillard2009} estimated an
upper limit of the mass surface density of \ion{H}{1} gas in the shocked
region as $5\times10^5~ M_\sun$~kpc$^{-2}$. Thus, the gas-to-dust mass
ratio is 170-210, which is in agreement with that of our
Galaxy~\citep{sodroski1997}. The estimated properties of the dust in 
the SQ are summarized in Table~\ref{prop}.

\section{Discussion}
\subsection{Shock-powered [\ion{C}{2}]158\,$\mu$m line
  emission from the shocked region}
To investigate the possibility of the luminous [\ion{C}{2}] line emission
from the shocked region, the C$^+$ abundance per hydrogen atom
$X_\mathrm{C^+}$ is estimated with assumptions that the [\ion{C}{2}] line
emission comes from the warm H$_2$ gas and is optically thin. Within
these assumptions, $X_\mathrm{C^+}$ is given as
  \begin{equation}
 X_\mathrm{C^+} = 1.1\times10^{-4}\left(
				    \frac{\Sigma_{L_{\mathrm{[CII]}}}\
				    L_\sun\mathrm{kpc}^{-2}}{\Sigma_{\mathrm{M_{H_2}}}\
				    M_\sun\mathrm{kpc}^{-2}}\right) 
 \left(\frac{1+2\mathrm{exp}(-91\
  \mathrm{K}/T)+n^\mathrm{crit}_\mathrm{H}/n_\mathrm{H}}{2\mathrm{exp}(-91\ \mathrm{K}/T)}\right),   
\label{eq7}
\end{equation}
where $T$ is the kinetic temperature of H$_2$ gas,
$n^\mathrm{crit}_\mathrm{H}$ is the critical density of the
[\ion{C}{2}] line emission for \ion{H}{1} collisions, and $n_\mathrm{H}$ is the
number density of \ion{H}{1}. At the shocked region, $T$ and
$n_\mathrm{H}$ are 158 K and $>10^3$ cm$^{-3}$,
respectively~\citep{culver2010}. $n^\mathrm{crit}_\mathrm{H}$ is 
$3.0\times10^3$ cm$^{-3}$ at 158 K~\citep{langer2010}. Thus, $X_\mathrm{C^+}$
is estimated to be $\sim1\times10^{-4}$. Assuming that the carbon in
the [CII] line emitting region is in singly ionized form, the carbon abundance
in the shocked region is in agreement with that in an interstellar gas-phase
($1.4\times10^{-4}$,~\cite{cardelli1996}). 
Therefore, the luminous [\ion{C}{2}] line emission from the shocked
region is physically plausible provided that C$^+$ is the main carbon
form in the warm H$_2$ gas.  

\subsection{Origin of the cold dust emission from the shocked region}
\begin{figure}
\begin{center}
 \includegraphics[angle=0,scale=0.70]{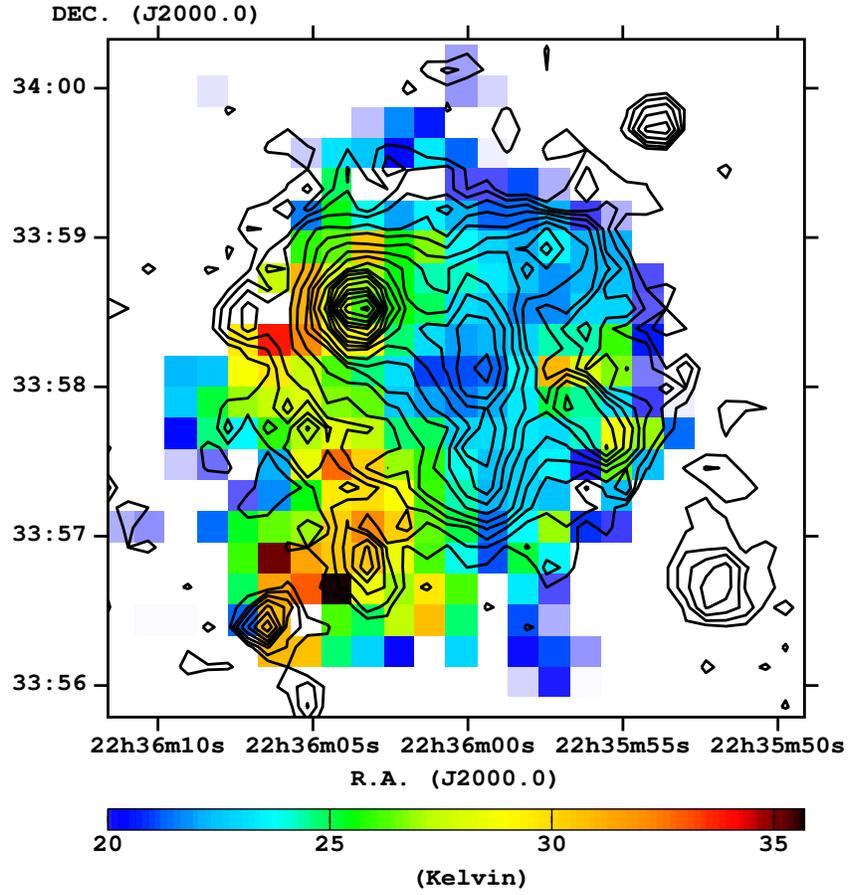}
\end{center}
 \caption{Color
 temperature map of the SQ. The contours show XMM/Newton
 X-ray contours that are the same as those in Fig.~\ref{4image}(f).} 
 \label{color}
\end{figure}

The clear spatial correlations among the cold dust, H$_2$, and X-ray
emissions suggest two possibilities: radiative heating of dust grains in
a low interstellar radiation field (ISRF) and collisional heating of
dust grains in a hot plasma. 
To explain the formation of H$_2$ gas in the shocked region, 
\cite{guillard2009} introduced a model of the shock in an
inhomogeneous gas phase. 
In their model, there are two critical assumptions to explain the formation
of H$_2$ gas in the collision age: dust survival in the shock and the Galactic
gas-to-dust mass ratio. The spatial correlation
between the maps of the cold dust emission and H$_2$ rotational
line strongly supports H$_2$ formation on dust grains. In addition,
the observed gas-to-dust mass ratio in the shocked region is in agreement
with the Galactic one. These facts indicate that the
H$_2$ formation time scale should be shorter than the collision age. In
the case of the multiphase medium, 
the cold dust emission comes from molecular gas  
clouds. \cite{guillard2010} modeled SEDs of dust emission associated
with diffuse (the hydrogen column density $N_\mathrm{H}\sim2\times10^{20}$ cm$^{-2}$) or clumpy
molecular gas ($N_\mathrm{H}\sim7\times10^{21}$ cm$^{-2}$) in radiation
fields. As discussed in Section 4.1, the [\ion{C}{2}] line emission can
come from molecular gas clouds. The optical depth of the [\ion{C}{2}]
line $\tau_\mathrm{[CII]}$ is expressed in Eq.~(A4) of
\cite{crawford1985}. By using $X_\mathrm{C^+}$ of $1\times10^{-4}$ and
the velocity dispersion of 870 km s$^{-1}$~\citep{culver2010}, $N_\mathrm{H}$ is 
estimated to be $2\times10^{24}$ cm$^{-2}$ when $\tau_\mathrm{[CII]}$
becomes unity. Thus, the [\ion{C}{2}] line emission is optically thin
for both cloud models. To estimate color temperatures of cold dust for
both cloud models, flux densities at 90 and 160 $\mu$m 
are obtained from Fig.~7 in \cite{guillard2009}. Assuming the modified
blackbody spectrum with the emissivity power-law index of unity, the
color temperature is estimated to be 24 K for the diffuse molecular gas
model and 20 K for the clumpy one. We investigate the spatial
distribution of the color temperature in the SQ (see Fig.~\ref{color}) by
using the flux densities at 90 and 160 $\mu$m that are above the
$3\sigma$ sky background noise. Errors in the color temperature are estimated
as $2.0$-$2.5$ K. The color temperature in the shocked region is about
20 K, which favors the temperature for the clumpy molecular gas model
rather than the diffuse one and significantly lower than that in the
surrounding galaxies. Thus, it is possible that cold dust in the shocked   
region is radiatively heated by the ISRF.

The spatial correlation between the maps of the cold dust emission and
X-ray suggests an alternative possibility for the 
origin of cold dust emission in the shocked region. 
If dust grains are exposed to the hot plasma for the collision age
($\sim 5\times 10^{6}$ yr), grains smaller than 0.1 $\mu$m in radius must have been
destroyed in $\sim 5\times 10^{6}$ yr~\citep{guillard2009}. We calculated the temperatures
of dust grains with radii $a>0.1$ $\mu$m heated collisionally by ambient
plasma electrons, based on the model of \cite{dwek1986,
dwek1987}. We assumed the size distribution $\mbox{d}n/\mbox{d}a \propto a^{-\alpha}$
with $\alpha$ = 2.5-3.5 over the range $0.1 < a < 1\ \mu  
\mbox{m}$ for either silicate or graphite grains. The density and
temperature of the electrons are taken to be $n_e=0.02$ cm$^{-3}$ and
$T_e=6 \times 10^6$ K, respectively~\citep{trinchieri2003}. As a result,
we derive the dust temperature of 20-22 K, which is insensitive to a
specific choice of $\alpha$ or grain composition. The derived
temperature is in a range similar to that observed in the shocked 
region. Thus we cannot distinguish between radiative heating and
collisional heating of dust grains solely from the dust temperature. 

However, the similarity in the spatial distribution between the
far-IR (cold dust and [\ion{C}{2}] line) emission and H$_2$ emission strongly
suggests that H$_2$ molecules form on dust grains and coexists with C$^{+}$. 
Moreover, \cite{natale2010} showed that the collisionally heated dust
emission should be minor contribution to explain the X-ray temperature
map. Therefore, it is likely that the far-IR dust emission
arises mostly from radiative heating of cold dust in clumpy molecular
gas. These facts support the multiphase medium model for the SQ.

\section{Summary}
We observed the SQ with the FIS aboard \textit{AKARI} in four far-IR
bands. The N160 band image shows single
peak emission in addition to the structure extending in the North-South
direction along the shock ridge as seen in the WIDE-L band, H$_2$
emission and X-ray emission. Whereas most of the far-IR emission in the
shocked region comes from the cold dust component,
the [\ion{C}{2}] line emission whose luminosity is comparable to that of
the warm H$_2$ gas can significantly contribute to the
single peak emission in the N160 band. It can be explained that the
[\ion{C}{2}] line emission comes from the warm H$_2$ gas.   
In the shocked region, the observed gas-to-dust mass ratio is in
agreement with the Galactic one. These indicate that H$_2$ molecules are
produced on the grain surface and their formation timescale should be
shorter than the collision age. The color temperature in the shocked
region ($\sim 20$ K) is significantly lower than that in the surrounding
galaxies ($\sim 30$ K). It is considered that the cold dust emission
comes mostly from radiative heating in dense and clumpy gas clouds.
Our results support the scenario of the multiphase medium of the
pre-shocked gas in the IGM.

\acknowledgments
We are grateful to the referee for providing us very important comments
and corrections on our calculation. The present work is based on
observations with \textit{AKARI}, a JAXA project with the participation
of ESA. The Digitized Sky Surveys were produced at the Space Telescope
Science Institute under U.S. Government grant NAG W-2166. The images of
these surveys are based on photographic data obtained using the Oschin
Schmidt Telescope on Palomar Mountain and the UK Schmidt Telescope. The
plates were processed into the present compressed digital form with the
permission of these institutions. This work was supported in part by
Grant-in-Aid for Young Scientists by MEXT (21740139).

\clearpage

\begin{table}
\vspace{10mm}
  \caption{Properties of the far-IR dust emission at the shocked region}\label{prop}
 \vspace{-2mm}
 \begin{center}
  \begin{tabular}{cccc}   
    \tableline \tableline 
    $T_\mathrm{C}$ & $\Sigma_{L_\mathrm{FIR}}$ & $\Sigma_{M_\mathrm{d}}$ &
   Gas-to-dust  \\ 
   (K) & (erg sec$^{-1}$ kpc$^{-2}$) & ($M_\sun$~kpc$^{-2}$) & mass ratio \\
   \tableline
   22$\pm$1 & $(2.0 \pm 0.6)\times 10^{40}$ &   $(1.1\pm0.3)\times10^4$ & 170$-$210 \\ \tableline 
  \end{tabular} 
 \end{center}
\end{table}


\begin{thebibliography}{}
\bibitem[Allen and Hartsuiker~(1972)]{allen1972}
Allen, R.~J., \& Hartsuiker, J.~W. 1972, \nat, 239, 324
\bibitem[Appleton et al.~(2006)]{appleton}
Appleton, P.~N. et al. 2006, \apjl, 639, L51

\bibitem[Boselli et al.~(2002)]{boselli2002}
Boselli, A., Gavazzi, G., Lequeux, J.,\& Pierini, D. 2002, \aap, 385, 454 

\bibitem[Cardelli et al.~(1996)]{cardelli1996}
Cardelli, J.~A., Meyer, D.~M., Jura, M., \& Savage, B.~D. 1996, \apj,
				     467, 334
\bibitem[Cluver et al.~(2010)]{culver2010}
Cluver, M.~E. et al. 2010, \apj, 710, 248

\bibitem[Crawford et al.~(1985)]{crawford1985}
Crawford, M.~K., Genzel, R., Townes, C.~H., Watson, D.~M. 1985, \apj,
				     291, 755
\bibitem[Dwek~(1987)]{dwek1987}
Dwek, E. 1987, \apj, 322, 812
\bibitem[Dwek~(1986)]{dwek1986}
Dwek, E. 1986, \apj, 302, 363
\bibitem[Gould \& Salpeter~(1963)]{gould1963}
Gould R.~J., Salpeter E.~E. 1963, \apj, 138, 393
\bibitem[Guillard et al.~(2010)]{guillard2010} 
Guillard, P., Boulanger, F., Cluver, M.~E., Appleton, P.~N., Pineau des
			     Forets, G.,\& Ogle, P. 2010, \aap, 518, 59
\bibitem[Guillard~(2010)]{guillard2010b}
Guillard, P. 2010, arXiv:1001.3613

\bibitem[Guillard et al.~(2009)]{guillard2009}
Guillard, P., Boulanger, F., Pineau Des For{\^e}ts, G., \& Appleton,
			     P.~N. 2009, \aap, 502, 515

\bibitem[Hickson~(1992)]{hickson1992}
Hickson, P., Mendes de Oliveira, C., Huchra, J.~P., \& Palumbo,
			     G.~G. 1992, \apj, 399, 353 

\bibitem[Hickson~(1982)]{hickson}
Hickson, P. 1982, \apj, 255, 382

\bibitem[Hildebrand(1983)]{hildebrand}
Hildebrand,~R.~H. 1983, \qjras, 24, 267.






\bibitem[Kaneda et al.~(2009)]{kaneda2009}
Kaneda, H., Koo, B.~C., Onaka, T., \& Takahashi, H.
2009, Advances in Space Research, 44, 1038
\bibitem[Kawada et al.~(2007)]{kawada}
Kawada, M. et al. 2007, \pasj, 59, 389



\bibitem[Langer et al.~(2010)]{langer2010}
Langer, W.~D., Velusamy, T., Pineda, J.~L., Goldsmith, P.~F., Li, D., \&
				     Yorke, H.~W. 2010, \aap, 521L, 17L

\bibitem[Moles et al.~(1997)]{moles1997}
Moles, M., Sulentic, J.~W., \& Marquez, I. 1997, \apjl, 485, L69

\bibitem[Murakami et al.~(2007)]{murakami}
Murakami, H. et al. 2007, \pasj, 59, 369

\bibitem[Natale et al.~(2010)]{natale2010}
Natale, G.~et al. 2010, \apj, 725, 955

\bibitem[O'Sullivan et al.~(2009)]{osullivan}
O'Sullivan, E., Giacintucci, S., Vrtilek, J.~M., Raychaudhury, S., \&
			     David, L.~P. 2009, \apj, 701, 1560
\bibitem[Pietsch et al.~(1997)]{pietsch}
Pietsch, W., Trinchieri, G., Arp, H., \& Sulentic, J.~W. 1997, \aap,
			     322, 89
\bibitem[Shirahata et al.~(2009)]{shirahata}
Shirahata, M. et al. 2009, \pasj, 61, 737


\bibitem[Sodroski et al.~(1997)]{sodroski1997}
Sodroski, T.~J., Odegard, N., Arendt, R.~G., Dwek, E., Weiland, J.~L.,
			     Hauser, M.~G. \& Kelsall, T. 1997, \apj,
			     480, 173S



\bibitem[Trinchieri et al.~(2005)]{trinchieri2005}
Trinchieri, G., Sulentic, J., Pietsch, W., \& Breitschwerdt, D. 2005,
			     \aap, 444, 697

\bibitem[Trinchieri et al.~(2003)]{trinchieri2003}
Trinchieri, G., Sulentic, J., Breitschwerdt, D., \& Pietsch, W. 2003,
			     \aap, 401, 173


\bibitem[Xu et al.~(2003)]{xu2003}
Xu, C.~K., Lu, N., Condon, J.~J., Dopita, M., \& Tuffs, R.~J. 2003,
			     \apj, 595, 665


\end{thebibliography}
\end{document}